\documentclass{article}
\usepackage{spconf,amsmath,graphicx}
\usepackage{algorithm}
\usepackage{algorithmicx}
\usepackage{algpseudocode}
\usepackage{amssymb}
\usepackage{hyperref}
\usepackage{cleveref}
\usepackage{makecell}
\usepackage{tabularx}
\usepackage{booktabs}
\usepackage{siunitx}

\title{Mandarin Lombard Flavor Classification}
\name{Qingmu Liu$^1$,Yuhong Yang$^{1,2*}$,Baifeng Li$^1$, Hongyang Chen$^1$, Weiping Tu$^1$, Song Lin$^3$}
\address{{$^1$ National Engineering Research Center for Multimedia Software, School of Computer Science,}\\
{Wuhan University, Wuhan, China.}\\
{$^2$ Hubei Key Laboratory of Multimedia and Network Communication Engineering, Wuhan, China.}\\
{$^3$ Guangdong OPPO Mobile Telecommunications Corp., China.}}

\begin{document}
%
\maketitle
\renewcommand{\thefootnote}{\fnsymbol{footnote}}
\footnotetext[1]{Correspondence: yangyuhong@whu.edu.cn } 

\begin{abstract}
The Lombard effect refers to individuals' unconscious modulation of vocal effort in response to variations in the ambient noise levels, intending to enhance speech intelligibility.
The impact of different decibel levels and types of background noise on Lombard effects remains unclear. Building upon the characteristic of Lombard speech that individuals adjust their speech to improve intelligibility dynamically based on the self-feedback speech, we propose a flavor classification approach for the Lombard effect. We first collected Mandarin Lombard speech under different noise conditions, then simulated self-feedback speech, and ultimately conducted the statistical test on the word correct rate. We found that both SSN and babble noise types result in four distinct categories of Mandarin Lombard speech in the range of 30 to 80 dBA with different transition points.
\end{abstract}
\section{Introduction}
\label{sec:intro}

The Lombard effect is a  vocal response whereby speakers involuntarily adjust their vocal effort in a noisy environment according to their self-auditory monitoring to make their speech more intelligible \cite{lombard1911signe,brumm2011evolution}.
Current Lombard datasets have background noise levels ranging from 35 to 96 dB, and have various types of noise, including pink noise, speech-shaped noise (SSN), car noise, etc. \cite{tartter1993some,bovril2011ut,hansen2009analysis}. However, it's uncertain how these noise levels and types relate to triggering the distinctive Lombard flavor. Prolonged speech in noisy environments may potentially result in hearing impairment and speech-related issues. Consequently, it is essential to determine the distinctive Lombard flavor based on the background noise level and type to offer guidance for the collection of Lombard datasets, ultimately aiming to minimize the workload associated with the acquisition of such datasets.

Previous studies on the Lombard effect aimed to identify the starting point of the Lombard effect, which is the transition from plain speech to Lombard speech, i.e., the first Lombard style. However, these findings lack consistency.

Pearsons et al. stated that teachers in the classroom environment began to raise their voice sound pressure level (SPL) when the background noise's SPL was above 48 dB \cite{pearsons1977speech}.

Yiu et al. recorded 24 vocally healthy young adults' 3 to 5 minutes monologue passage in the clinic room (35.5 dBA), clinic corridor (54.5 dBA), and pantry room (67.5 dBA) situations. 
Speakers' voice SPL showed significant increases in the  67.5 dBA environment than in the other two conditions. They deduced that background noise adversely affected voice production when the noise level was beyond 60 dBA \cite{yiu2016effect}.

Pasquale Bottalico investigated the starting point of the Lombard effect in relation to background noise decibel levels. twenty subjects, consisting of men and women aged 18 to 34, were recruited to carry out recordings in an anechoic chamber. They employed 10 decibel levels of pink noise, ranging from 20 dBA to 65 dBA with 5 dBA increments. To emulate a realistic conversation, a listener was positioned 2.5 meters away from the speaker, who read six sentences from the Rainbow passage. The vocal effort of speakers was evaluated using sound pressure levels. Then they fit change curves at various decibel levels and identify the point where the slopes altered. The study concluded that the Lombard effect starts at 43.3 dBA, based on a noticeable change in the speaker's vocal effort curve at that noise level \cite{bottalico2017evaluation}. 

Then Bottalico conducted a similar experiment with typical restaurant noise between 35 and 85 dBA in 2018. However, the starting point moved to 57.3 dBA. Bottalico speculated that the difference could be due to that automatic mechanism in speech regulation related to privacy \cite{bottalico2018lombard}.

One possible explanation for the divergent conclusions reached by existing studies is that they use speakers' voice SPL to classify. Research shows that even at the same SPL, Lombard speech exhibits higher intelligibility than ordinary speech \cite{summers1988effects}. Thus, it would be more reasonable to evaluate the variability of the Lombard effect using speech intelligibility. Another potential explanation for the differences in past research findings may stem from the use of varied noise types like classroom, crowd, and pink noise, potentially eliciting distinct Lombard effect styles.

Further, Hansen et al. first proposed that the Lombard effect has a different style \cite{hansen2009analysis}. They used a Gaussian Mixture Model (GMM) trained with 19 female speakers' speech to classify 3-s duration Lombard utterances' noise level (from 65 to 90dB SPL, 5 dB increments) and noise type (car noise, large crowd noise, and pink noise).  The classification performance is significantly different from random. Thus, they suggested that Lombard speech exists distinct flavors responding to varying noise levels and types. 

Although Hansen proposed that the Lombard effect under different background noise types belong to different Lombard flavors, in past studies, it was generally agreed that the type of noise had no significant effect on the Lombard effect. 

T Letowski et al. compared vocal pitch and overall SPL of Lombard speech produced by 10 subjects(5 men and 5 women) in multi-talker, traffic, and wideband noise presented at 70 and 90 dB SPL. No significant differences exist between noise types, indicating that the type of noise has no specific effect on the Lombard effect \cite{letowski1993acoustical}. Maria Södersten also claimed no significant differences in Lombard speech's SPL between soft noise and day-care babble, and no significant differences were observed between disco and loud noise \cite{sodersten2005loud}.

Studies have found inconsistent and even contradictory results regarding the effect of noise levels and noise types on the Lombard flavor. Therefore, it is necessary to develop an approach that can classify the Lombard effect under varying noise levels.

In our previous study, we attempted to analyze the Lombard flavor by combining the self-feedback speech with different noise levels and then using statistical tests on the short-time objective intelligibility (STOI) 
 \cite{taal2010short}scores of the mixed audio \cite{yang22i_interspeech}. 
However, the study limited its focus to Lombard classification under SSN, neglecting other noise types. The use of nonsensical Grid sentences likely weakened Lombard effect elicitation \cite{tasko2004variations}. Additionally, the absence of the STOI-to-word correct rate (WCR) mapping questions the reliability of the results.

Based on this work, we made the following improvements:
(1)in contrast to using nonsensical sentences, which have been shown to induce a less Lombard effect \cite{lbf}, we selected sentences with high naturalness, longer length, and covering the maximum number of Mandarin tones and tonal combinations to elicit a more pronounced Lombard effect;  (2) we introduced babble noise to verify the existence of different Lombard flavors under varying noise types; (3) the STOI scores were mapped to subjectively measured word correct rates to enhance the reliability of the statistical test.
 
We observed that Mandarin Lombard speech exhibits four distinct patterns under both SSN and babble noise in the 30-80 dBA range. The transition points differ: 45, 65, and 75 dBA for SSN and 55, 65, and 75 dBA for babble noise.
\begin{table*}[th]
\caption{Results of the classification of Lombard flavor over SSN levels. The "t" columns represent the statistical test results between two noise conditions. Symbols: increase $\uparrow$, decrease $\downarrow$.   $\ast$  indicates a significant difference (p-value\textless0.001).}
  \label{tab:ssn_level}
  \centering
  \scalebox{1}{
  \begin{tabular}{cccccccc}
\toprule
 \multicolumn{1}{c}{\bfseries{\makecell[c]{dBA level \\ combination}}} &
 \multicolumn{1}{c}{\bfseries{\makecell[c]{lower level}}} &
 \multicolumn{1}{c}{\bfseries{\makecell[c]{higher level}}} &
 \multicolumn{1}{c}{\bfseries{\makecell[c]{t}}} &
\multicolumn{1}{c}{\bfseries{\makecell[c]{dBA level \\ combination}}} &
 \multicolumn{1}{c}{\bfseries{\makecell[c]{lower level}}} &
 \multicolumn{1}{c}{\bfseries{\makecell[c]{higher level}}} &
 \multicolumn{1}{c}{\bfseries{\makecell[c]{t}}} \\
\midrule
$30/35$  & $97.38\pm0.44$&  $97.66\pm0.45$  &$\uparrow$  & $45/60$& $91.26\pm1.49$&  $92.58\pm1.68$  &$\uparrow$\\

$30/40$  & $94.61\pm1.24$&  $95.59\pm1.26$  &$\uparrow$  & $\pmb{45/65}$    & $\pmb{82.07\pm2.44}$&  $\pmb{85.32\pm2.00}$  &$\pmb{ \  \uparrow^ *}$       \\

$\pmb{30/45}$  & $\pmb{89.56\pm2.30}$&  $\pmb{92.27\pm1.33}$  &$\pmb{\  \uparrow^ *}$  & $65/70$    & $81.59\pm2.92$&  $83.28\pm3.28$  &$\uparrow$       \\

$45/50$  & $92.20\pm1.41$&  $92.12\pm1.63$  &$\downarrow$  & $\pmb{65/75}$    & $\pmb{76.40\pm3.16}$&  $\pmb{79.82\pm3.46}$  &$\pmb{\  \uparrow^ *}$       \\

$45/55$  & $93.23\pm1.19$&  $93.53\pm1.20$  &$\uparrow$  & $75/80$    & $63.49\pm6.33$&  $62.50\pm6.96$  &$\downarrow$       \\

    \bottomrule
    
\end{tabular}}
  
\end{table*}
\section{Lombard Speech dataset collection}
\label{sec:classification}

\subsection{Subjects Selection}

To mitigate the impact of inter-speaker variability, the participant pool was expanded from 4 to 10. Participants were 10 students, 5 males and 5 females, in the 20-23 year age range. All participants reported no hearing impairment and scored Class 2 Level 1 on the Putonghua Shuiping Ceshi proficiency test. They were all paid for their participation. 

A simple online hearing test was conducted on the Philips HearLink hearing aid official website before the recording began.  After the test showed that the participants had no hearing loss, we would move to the recording step.

\begin{figure}[htb]
  \centering
  \includegraphics[height=0.18\textwidth,width=0.72\linewidth]{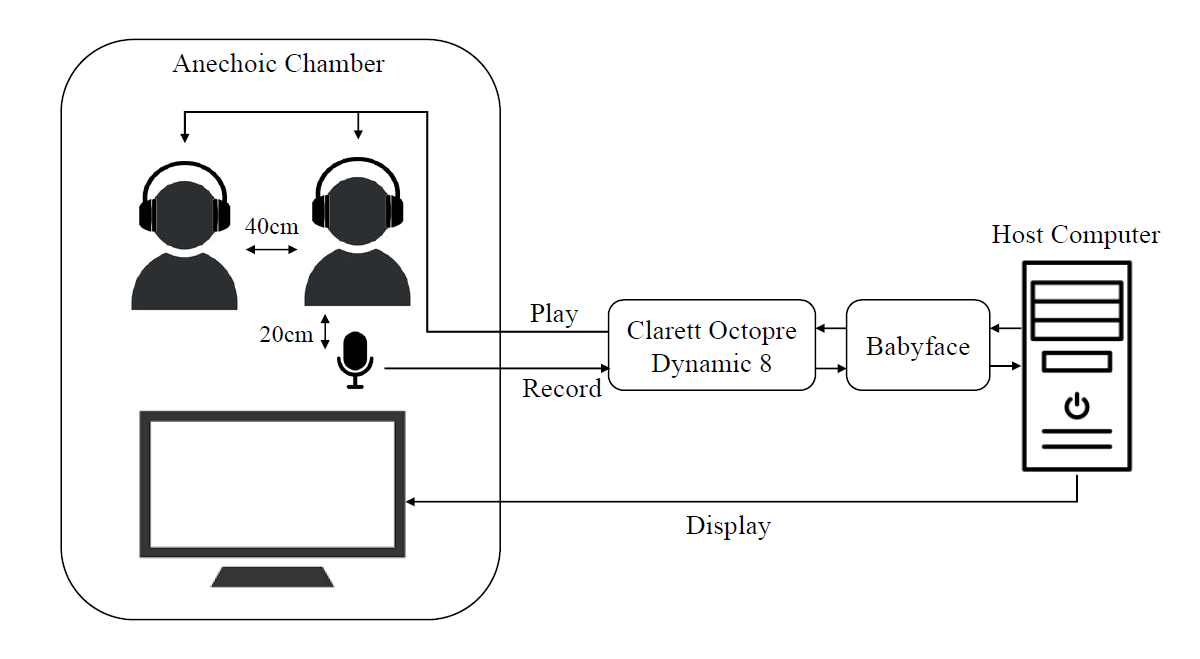}
  \caption{Two participants sat 40 cm apart in an anechoic chamber, wearing headphones. A microphone was placed 20 cm from the speaker. Automated recording software appeared on a nearby touchscreen. The microphones and headphones were connected to the Focusrite Clarett OctoPre Dynamic 8, which was further connected to the Babyface sound card.}
  \label{fig:record}
\end{figure}
\subsection{Sentence Material}
Sentences were chosen from the Global TIMIT Mandarin Chinese "Calibration" set, a total of 20 sentences, covering the maximum number of tones and tonal combinations. Every subject read the same 20 sentences under each noise condition being tested. 

\subsection{Background Noise} 

Research showed that the Lombard effect induced by background noise whose spectral composition is similar to speech noise is more pronounced than that evoked by background noise without speech frequency components \cite{garnier2010influence,stowe2013evidence}. 

SSN is stable, which provides greater experimental control, and has been shown to elicit the Lombard effect more effectively. Thus, we chose SSN as the background noise to classify Lombard styles under different decibel levels of noise. 
We also included babble noise to simulate real-world auditory environments. Its complexity allows for a nuanced exploration of the Lombard effect under conditions that more closely mimic natural human interactions.

\subsection{Recording Conditions and Procedures}

Lombard speech was recorded within the anechoic chamber at Wuhan University to ensure optimal acoustic conditions. The recording scenario is shown in Figure ~\ref{fig:record}.

To simulate a realistic communication scenario, two participants sat \SI{40}{\centi\metre} apart, facing a touch screen with recording software. Both participants wore Sennheiser HD 300Pro closed-back headphones, while a RODE NT1-A microphone was situated \SI{15}{\centi\metre} from the speaker. The audio files were recorded at a sample rate of 48 kHz.

The closed-back headphones presented various noise conditions in a randomized order. To guarantee that the noise levels experienced by the speakers through the headphones were equivalent to those in an open field, we calibrated the headphones accordingly. As closed headphones attenuate air-conducted speech, we use TotalMix FX to compensate. This allows speakers to hear their own voices through the headphones, ensuring a comparable perception of speech with and without the headphones.

During the recording process, one participant read the displayed text while the other acted as the listener to confirm its accuracy before proceeding. Speakers were permitted to take breaks and rehydrate as needed throughout the recording session.

\section{Lombard flavor classification}
\subsection{Self-feedback Speech Model}

A key component of the Lombard effect is self-monitoring, which uses auditory feedback to detect speech errors \cite{garnier2010influence}. Thus, generating self-feedback speech to add bone-conducted speech can better simulate the process of Lombard speech generation.

To simulate the speech that speakers heard by themselves, we employed a two-parameter self-feedback voice model involving both air-conducted and bone-conducted pathways \cite{won2014simulation}.
We obtained self-feedback speech by applying the transformation function to the recorded speech captured by the microphone.

\subsection{Mapping Objective Intelligibility Scores to Subjective Word Correct Rate} 

Studies have demonstrated that Lombard speech is more intelligible than plain speech (uttered in quiet) when presented at the same signal-to-noise ratio (SNR) 
 \cite{summers1988effects,cooke2014contribution}.  Therefore, using speech intelligibility to classify the Lombard style is reasonable. We used the widely-accepted STOI measure \cite{taal2010short}  to assess intelligibility. Drawing from that, we designed intelligibility mapping experiments for speech samples with both SSN and babble background noises.

We curated a dataset from 34 speakers, each recording 100 sentences at noise levels of 40dBA and 80dBA \cite{lbf}. Randomly selected sentences from the 40dBA set were processed through three normal to Lombard voice conversion models, namely LSTM \cite{liganglstm}, CycleGAN \cite{cyclegan} , and StarGAN \cite{ligangStarGAN}, creating three conditions' speech. Additional conditions were formed using random samples from the original 40dBA and 80dBA sets. The evaluation was conducted at ten distinct SNR levels, ranging from -18.5 to 4dB, using both SSN and babble noises.

For the evaluation phase, 15 subjects participated in a speech audiometry hearing test. Each subject was exposed to two sentences under each condition at every SNR level. Participants transcribed sentences heard under various conditions and SNRs, and the word correct rate (WCR) was calculated as the intelligibility metric.

Finally, STOI scores corresponding to these audio samples were calculated and then mapped to subjective intelligibility metrics using the mapping function
 $$f(d)=\frac{100}{1+\exp (a d+b)}$$
thereby accomplishing the intended nonlinear translation between the objective and subjective measures of speech intelligibility.

\begin{figure}[htb]
  \centering
  \includegraphics[height= 0.18\textwidth,width=0.72\linewidth]{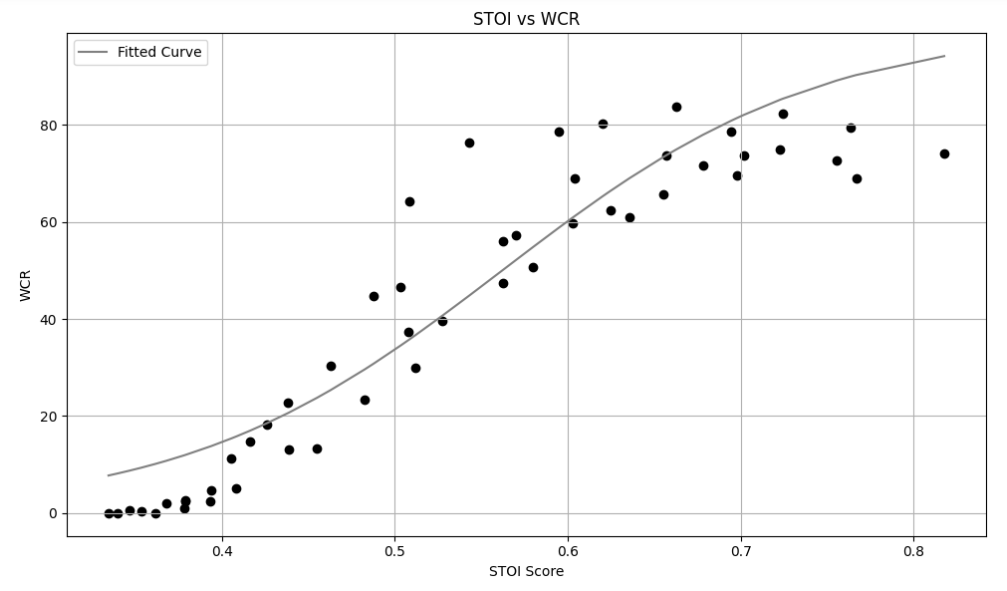}
  \caption{ The gray line denotes the mapping used to translate the objective output to an intelligibility score.}
  \label{mapping}
\end{figure}

\begin{table*}[th]
\caption{Results of the classification of Lombard flavor over babble noise levels. The "t" columns represent the statistical test results between two noise conditions. Symbols: increase $\uparrow$, decrease $\downarrow$.   $\ast$  indicates a significant difference (p-value\textless0.001).}
  \label{tab:babble_level}
  \centering
  \scalebox{1}{
  \begin{tabular}{cccccccc}
\toprule
 \multicolumn{1}{c}{\bfseries{\makecell[c]{dBA level \\ combination}}} &
 \multicolumn{1}{c}{\bfseries{\makecell[c]{lower level}}} &
 \multicolumn{1}{c}{\bfseries{\makecell[c]{higher level}}} &
 \multicolumn{1}{c}{\bfseries{\makecell[c]{t}}} &
\multicolumn{1}{c}{\bfseries{\makecell[c]{dBA level \\ combination}}} &
 \multicolumn{1}{c}{\bfseries{\makecell[c]{lower level}}} &
 \multicolumn{1}{c}{\bfseries{\makecell[c]{higher level}}} &
 \multicolumn{1}{c}{\bfseries{\makecell[c]{t}}} \\
\midrule
$30/35$  & $96.69\pm0.83$&  $96.97\pm0.58$  &$\uparrow$  & $55/60$& $90.97\pm1.94$&  $92.14\pm1.92$  &$\uparrow$\\

$30/40$  & $93.97\pm1.74$&  $94.09\pm1.61$  &$\uparrow$  & $\pmb{55/65}$    & $\pmb{77.49\pm3.48}$&  $\pmb{81.85\pm3.02}$  &$\pmb{ \  \uparrow^ *}$       \\

$30/45$  & $89.40\pm2.75$&  $90.63\pm1.83$  &$\uparrow$  & $65/70$    & $77.50\pm4.155$&  $76.87\pm4.35$  &$\downarrow$       \\

$30/50$  & $89.06\pm2.63$&  $91.38\pm2.06$  &$\uparrow$  & $\pmb{65/75}$    & $\pmb{68.60\pm4.70}$&  $\pmb{70.94\pm5.14}$  &$\pmb{\  \uparrow^ *}$       \\

$\pmb{30/55}$  & $\pmb{90.13\pm2.66}$&  $\pmb{93.38\pm1.57}$  &$\pmb{\  \uparrow^ *}$  & $75/80$    & $50.1\pm6.72$&  $51.1\pm6.18$  &$\uparrow$       \\

    \bottomrule
    
\end{tabular}}
  
\end{table*}

\subsection{Iterative Statistical Test}

Self-feedback speech at two decibel levels was overlaid with noise at the higher decibel level of each pair resulting in two groups of audio. We adjusted the energy of lower-decibel speech to match higher-decibel speech before overlaying. This ensures that the combined audio samples in two groups have the same SNR.

Vocal behavior in relation to noise exposure is highly individual \cite{lindstrom2011observations}, and gender differences were also noticed in the size of the Lombard effect \cite{alghamdi2018corpus}. In order to obtain a general conclusion, we calculated the average WCR for the same sentence's mixed audio (self-feedback speech overlaid with noise) from all subjects, resulting in 20 WCR for each group of such tests. Using an average WCR reduces variability in the data, thereby contributing to the overall rigor and validity.

We iterated the statistical test as follows: if no significant differences were observed, we then repeated the statistical test steps with a higher level of noise overlaid with the lower level's speech used in the current test; otherwise, we continued the iteration starting from the higher level of the current test.

\section{Results}
\subsection{Mapping Result}
The means of the root of the mean squared prediction error (RMSE) $\sigma$ and correlation coefficient $\rho$ of WCR and STOI before mapping are $51.1522$ and $0.9189$ respectively. 

The result of mapping STOI to WCR is shown in Figure~\ref{mapping}. The obtained values for the free parameters of the nonlinear mappings, denoted as  $a$ and $b$, are $-10.88$ and $6.12$ respectively. The RMSE $\sigma$ and correlation coefficient $\rho$ are improved to $11.0692$ and $0.9343$ respectively.

\subsection{Lombard Flavor Classification Result}
A total of 10 speakers (balanced by gender) were recruited to participate in creating the dataset, each recording Lombard speech across 11 distinct noise levels and 2 noise types. In each noise condition, the speakers recorded 20 sentences, resulting in a total of 4400 recorded sentences.

The results for Lombard flavor classification over decibel levels under SSN and babble noise are shown in Table~\ref{tab:ssn_level} and Table~\ref{tab:babble_level}.
Mandarin Lombard speech under SSN and babble noise both fall into 4 categories in the 30-80 dBA range. However, the SSN transition points are at 45, 65, and 75 dBA, while for babble they are at 55, 65, and 75 dBA. These findings suggest that the pressure level of noise significantly impacts the intelligibility of Lombard speech. The difference in transition points indicates that the type of noise impacts the Lombard effect in distinct ways.

\section{Conclusion and Discussion}

This paper aims to address the challenge of the ambiguous boundaries of Lombard speech and to provide guidance for the efficient acquisition of Mandarin Lombard speech datasets. 
The divergence in our noise level classification results from those presented in 
 \cite{yang22i_interspeech}may be attributed to a more pronounced Lombard effect observed in meaningful text \cite{lbf}. Furthermore, we expanded our participant pool to mitigate the impact of individual variations in the Lombard effect across different speakers.
Our research findings align with those of \cite{bottalico2017evaluation}, both suggesting that the onset of the Lombard effect occurs between 40-45 dBA. Additionally, we corroborate the views expressed in \cite{hansen2009analysis} regarding the existence of distinct Lombard flavors under varying noise levels and types.

\section{Acknowledgments}
This research is funded in part by the National Natural Science Foundation of China (62171326), Key Research and Development Program of Hubei Province (220171406) and Guangdong OPPO Mobile Telecommunications Corp.

\small
\bibliographystyle{IEEEbib}
\bibliography{strings,refs}

\end{document}